\documentclass[sigconf,natbib=true,anonymous=false]{acmart}
\usepackage[table]{xcolor}
\usepackage{amsfonts}
\usepackage{algorithm}
\usepackage{algorithmic}
\usepackage{newfloat}
\usepackage{listings}
\usepackage{color, xspace}
\usepackage{pifont}

\usepackage{tikz}
\usepackage{xcolor}
\usepackage{multirow}
\usepackage{makecell}
\usepackage{siunitx}        % 数值对齐
\usepackage{booktabs}       % 美观的表格线
\usepackage{threeparttable} % 添加表格注释
\usepackage[edges]{forest}
\AtBeginDocument{%
  }
\setcopyright{acmlicensed}
\copyrightyear{2018}
\acmYear{2018}
\acmDOI{XXXXXXX.XXXXXXX}
\acmConference[Conference acronym 'XX]{Make sure to enter the correct
  conference title from your rights confirmation email}{June 03--05,
  2018}{Woodstock, NY}
\acmISBN{978-1-4503-XXXX-X/2018/06}

\begin{document}

\title[MALLOC: Benchmarking the Memory-aware Long Sequence Compression for Large Sequential Recommendation]{MALLOC: Benchmarking the Memory-aware Long \\Sequence Compression for Large Sequential Recommendation}

% \settopmatter{authorsperrow=4}
\author{Qihang Yu}
\authornote{Equal contribution.}
\orcid{0009-0000-2650-5080}
\email{yuqihang@zju.edu.cn}
\affiliation{%
  \institution{Zhejiang University}
  \city{Hangzhou}
  \country{China}
}
\author{Kairui Fu}
\authornotemark[1]
\orcid{0009-0004-0284-671X}
\email{fukairui.fkr@zju.edu.cn}
\affiliation{%
  \institution{Zhejiang University}
  \city{Hangzhou}
  \country{China}
}
\author{Zhaocheng Du}
\authornotemark[1]
\orcid{0000-0002-1811-129X}
\email{zhaochengdu@huawei.com}
\affiliation{%
  \institution{Huawei Noah's Ark Lab}
  \city{Hangzhou}
  \country{China}
}
\author{Yuxuan Si}
% \orcid{}
\email{syx_sue@zju.edu.cn}
\affiliation{
  \institution{Zhejiang University}
  \city{Hangzhou}
  \country{China}
}
\author{Kaiyuan Li}
\orcid{0009-0003-5823-7027}
\email[]{likaiyua23@mails.tsinghua.edu.cn}
\affiliation{
  \institution{Tsinghua University}
  \city{Beijing}
  \country{China}
}
\author{Weihao Zhao}
% \orcid{}
\email{weihaozhao@zju.edu.cn}
\affiliation{
  \institution{Zhejiang University}
  \city{Hangzhou}
  \country{China}
}
\author{Zhicheng Zhang}
\orcid{0009-0002-5676-2461}
\email{zhang-zc24@mails.tsinghua.edu.cn}
\affiliation{
  \institution{Tsinghua University}
  \city{Beijing}
  \country{China}
}
\author{Jieming Zhu}
\orcid{0000-0002-5666-8320}
\email{jamie.zhu@huawei.com}
\affiliation{%
  \institution{Huawei Noah's Ark Lab}
  \city{Hangzhou}
  \country{China}
}
\author{Quanyu Dai}
\orcid{0000-0001-7578-2738}
\email{daiquanyu@huawei.com}
\affiliation{%
  \institution{Huawei Noah's Ark Lab}
  \city{Hangzhou}
  \country{China}
}
\author{Zhenhua Dong}
\orcid{0000-0002-2231-4663}
\email{dongzhenhua@huawei.com}
\affiliation{%
  \institution{Huawei Noah's Ark Lab}
  \city{Hangzhou}
  \country{China}
}
\author{Shengyu Zhang}
\authornotemark[2]
\orcid{0000-0002-0030-8289}
\email{sy_zhang@zju.edu.cn}
\affiliation{
  \institution{Zhejiang University}
  \city{Hangzhou}
  \country{China}
}
\author{Kun Kuang}
\authornotemark[2]
\orcid{0009-0000-7528-8131}
\email{kunkuang@zju.edu.cn}
\affiliation{
  \institution{Zhejiang University}
  \city{Hangzhou}
  \country{China}
}
\author{Fei Wu}
\authornote{Corresponding author.}
\email{wufei@zju.edu.cn}
\orcid{0000-0003-2139-8807}
\affiliation{
  \institution{Zhejiang University}
  \city{Hangzhou}
  \country{China}
}
% \affiliation{
%   \institution{Shanghai AI Laboratory}
%   \city{Shanghai}
%   \country{China}
% }

\renewcommand{\shortauthors}{Qihang Yu et al.}

\begin{abstract}
The scaling law, which indicates that model performance improves with increasing dataset and model capacity, has fueled a growing trend in expanding recommendation models in both industry and academia. However, the advent of large-scale recommenders also brings significantly higher computational costs, particularly under the long-sequence dependencies inherent in the user intent of recommendation systems. Current approaches often rely on pre-storing the intermediate states of the past behavior for each user, thereby reducing the quadratic re-computation cost for the following requests. Despite their effectiveness, these methods often treat memory merely as a medium for acceleration, without adequately considering the space overhead it introduces. This presents a critical challenge in real-world recommendation systems with billions of users, each of whom might initiate thousands of interactions and require massive memory for state storage. Fortunately, there have been several memory management strategies examined for compression in LLM, while most have not been evaluated on the recommendation task. To mitigate this gap, we introduce MALLOC, a comprehensive benchmark for memory-aware long sequence compression. MALLOC presents a comprehensive investigation and systematic classification of memory management techniques applicable to large sequential recommendations. These techniques are integrated into state-of-the-art recommenders, enabling a reproducible and accessible evaluation platform. Through extensive experiments across accuracy, efficiency, and complexity, we demonstrate the holistic reliability of MALLOC in advancing large-scale recommendation. Code is available at \url{https://anonymous.4open.science/r/MALLOC}.
\end{abstract}





\maketitle

\section{Introduction}
With the exponential growth of online data and user activity, recommender systems have become essential in delivering personalized content across various domains like \emph{E-commerce}~\cite{zhou2018deep,xiao2020deep,yu2025thinkrec,fu2025forge} and \emph{videos}~\cite{cai2023two,covington2016deep}. Traditional methods predominantly follow the \emph{Embedding} + \emph{MLP} schema, whose architecture is relatively small in parameter count, while the feature engineering rather than interaction sequences holds a dominant position to improve the overall performance. In contrast, recent insights from the scaling law~\cite{kaplan2020scaling}, which suggest that the performance follows a certain power-law improvement with the increase of data volume and parameters, introduce a completely paradigm shifting from \emph{feature-centric} to \emph{parameter-centric} long sequence recommendation~\cite{zhai2024actions}.

\begin{figure}[b]
\vspace{-1em}
  \centering
  \includegraphics[width=1.0\linewidth]{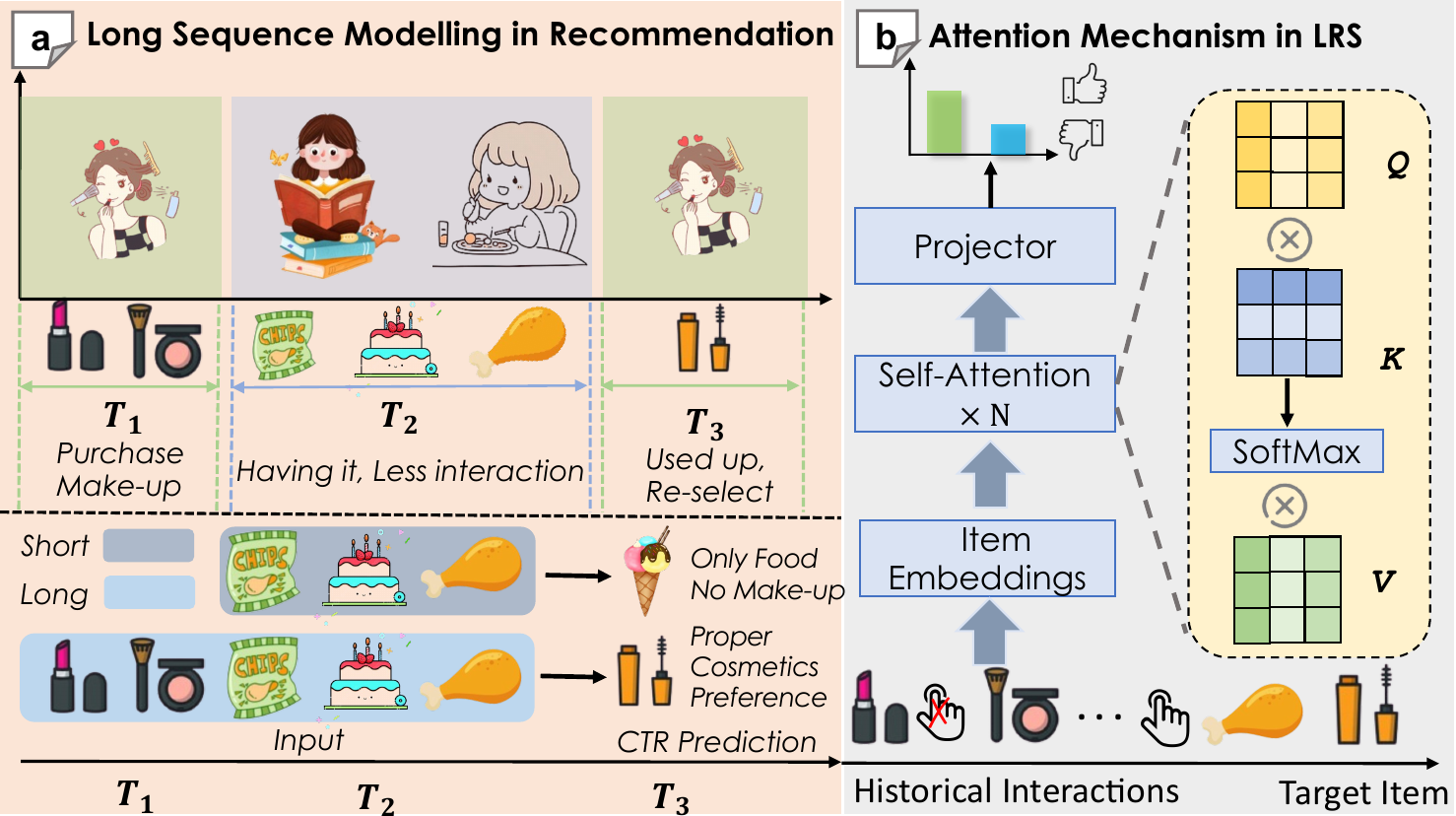}
  \caption{(a) Long-term historical interactions are crucial for capturing periodic user behaviors. (b) The standard self-attention mechanism in Large Recommender Systems.
  }
  \label{fig: intro}
\end{figure}
Despite the impressive performance of large sequential recommendation, the widely-used attention mechanism for sequence modeling~\cite{vaswani2017attention} in Figure~\ref{fig: intro}(b) might introduce significant computational overhead. This is primarily due to the quadratic cost of dot-product operations during inference, which scales with the square of the sequence length. Such inefficiency becomes a critical barrier in industrial deployment scenarios where long user interaction sequences ($\ge 1000$) are of great significance to recommendation diversity and effectiveness~\cite{chai2025longer,liao2025patchrec,pi2020search,si2024twin}. For example, as shown in Figure~\ref{fig: intro}(a), users may reduce engagement with makeup products after exhausting their current stock, while taking into account the historical interaction with cosmetics is crucial for their next purchase. Additionally, users often issue multiple requests within a single session~\cite{gong2020edgerec,qian2022intelligent}, requiring continuous responses from the system. Both scenarios highlight the urgent need for efficient inference to support scalable and practical deployment.
To quantify this bottleneck, our empirical study on a long-sequence model (seq len 1,024) reveals a critical "Memory–Latency Dilemma" at a batch size of 1,000: \textbf{Storage Explosion.} Full key-value (KV) caching demands ~545 GB of memory, far exceeding typical single-node GPU capacities and making resource provisioning prohibitive. \textbf{Computation Surge.} Conversely, re-computation spikes computational cost by ~450$\times$ (from 16 G to 7.1 T MACs), inevitably violating latency constraints. This dilemma renders both extremes unscalable, underscoring the urgency for memory-aware compression.

To retain all the raw-sequence information and enable end-to-end ranking~\cite{pi2020search,si2024twin}, current applications~\cite{chai2025longer,lv2024marm,zhai2024actions} commonly adopt memory-aware designs that retain pre-generated intermediate hidden states at different granularities. By reusing these states during inference, such designs avoid redundant computation and improve efficiency. While some of them have been deployed in real-world platforms~\cite{chai2025longer,lv2024marm}, they still have the following shortcomings that hinder the development of this field:
\begin{itemize}
\item \textbf{Limited Detail on Memory Structure Implementation.} All these methods provide limited detail about their memory construction. Currently, researchers~\cite{zhanglong,li2024snapkv} have verified that distinct implementations may affect the overall performance, making it challenging for researchers to discover the most effective one to implement. Till now, few works offer empirical evidence or experiments that could motivate practical deployments in recommendation, lacking the necessary support to inspire future research directions.
\item \textbf{Neglect of Inference Efficiency Comparison.} Efficient inference is essential for deploying large sequential recommenders in real-world settings and represents one of the core motivations behind using memory-based methods. The efficiency gains achieved through such techniques should, therefore, be rigorously evaluated and compared across different approaches. However, existing methods~\cite{lv2024marm,zhai2024actions,chai2025longer} typically only state that memory is used to accelerate the autoregressive process, without providing a systematic or fair comparison of their actual improvements.
\item \textbf{Lack of Memory Storage Consideration.} Even if memory mechanisms can improve inference speed, they also introduce significant storage demands by requiring the system to maintain intermediate states for all users during serving~\cite{chai2025longer}. The massive user base in recommendation systems imposes further challenges on the storage. An effective design should balance both recommendation accuracy and storage efficiency, yet the latter aspect lacks well-established guidelines for design and selection in current research.
\end{itemize}

To this end, we propose \textbf{MALLOC}, a comprehensive benchmark for memory-aware long-sequence compression for large sequential recommendation. To the best of our knowledge, MALLOC represents one of the first benchmarks explicitly designed for memory-aware compression in this setting, with a specific emphasis on preserving complete raw user interaction sequences in large recommender systems while simultaneously maintaining high inference efficiency.
We first conduct a systematic integration of existing long-sequence compression techniques from the perspective of memory allocation granularity, examining how intermediate states are stored, shared, or compressed at different levels during inference,
including several methods that have not previously been applied to or assessed within recommender systems. In addition to conventional recommendation performance metrics ($\Uparrow$), MALLOC introduces two central evaluation dimensions: computational overhead ($\Downarrow$, \textbf{Challenge \romannumeral2}) and memory occupation ($\Downarrow$, \textbf{Challenge \romannumeral3}), both of which are critical for the practical online deployment of large-scale recommender systems.
To facilitate fair and consistent comparisons, all baseline methods are implemented with unified data preprocessing procedures and harmonized model architectures, thereby minimizing confounding experimental variability. Moreover, we evaluate and compare the practical implementation complexity of each method, with the goal of informing method selection and system design in both academic research and industrial applications (\textbf{Challenge \romannumeral1}).

In summary, the main contributions of our paper are as follows:
\begin{itemize}
\item We introduce MALLOC, the first benchmark that systematically investigates long-sequence compression for large recommender systems from the perspective of memory allocation granularity to our knowledge. This structured framework unifies disparate long-sequence compression methodologies, offering a new perspective to understand how intermediate states are managed in large recommender systems.
\item We establish a rigorous multi-dimensional evaluation protocol that couples standard ranking metrics with system-level constraints. By constructing the accuracy-resource Pareto frontier, we expose the complex trade-offs between recommendation performance and deployment costs, addressing the critical gap in evaluating industrial feasibility.
\item Through extensive experiments on diverse datasets, we provide actionable deployment guidelines. We explicitly analyze the scalability of different architectures and their implementation complexity, identifying which memory-aware designs remain robust in deep networks and which offer the best return on engineering investment.
\end{itemize}

\section{Related Work}

\subsection{Large Recommender System}
Traditional recommendation systems have primarily focused on feature engineering with relatively compact architectures. However, recent advancements in large-scale recommendation have shifted the focus toward scaling laws, where expanding model capacity and representation space becomes a primary driver of performance improvement~\cite{zhang2024wukong,lv2024marm,zhai2024actions}.
For example, HSTU~\cite{zhai2024actions} redefines the recommendation task within a generative framework, employing stacked hierarchical sequential transduction units to strengthen feature interactions and selection. Building on this foundation, MTGR~\cite{han2025mtgr} bridges classical and modern paradigms by integrating widely used cross features from traditional models into its architecture. In contrast, MARM~\cite{lv2024marm} enhances a conventional recommender system~\cite{zhou2018deep} by adding self-attention~\cite{vaswani2017attention} layers prior to the final attention module, preserving architectural simplicity while expanding representational power.
Alongside architectural scaling, recent studies have begun to recognize that memory consumption, rather than parameter count alone, constitutes a major bottleneck in large-scale recommendation systems\cite{Petrov2024recjpq}.
These approaches collectively demonstrate a consistent trend: as model sizes grow to hundreds of billions of parameters, they achieve remarkable scalability and performance gains. However, the trade-offs between computational efficiency and model complexity remain critical challenges in real-world deployment.

\subsection{Long‑Sequence Recommendation}

In real-world applications, user behavioral histories often consist of thousands of interactions, posing significant challenges for efficient modeling. To address this, two primary solution strategies have emerged. First, retrieval-then-ranking two-stage approaches, such as SIM~\cite{pi2020search}, employ a coarse retrieval stage to identify relevant subsequences followed by a fine-grained ranking stage to refine recommendations. Similarly, TWIN~\cite{si2024twin} integrates dual-tower retrieval with attention-based ranking modules to achieve scalable performance. However, these methods typically process only a subset of the full interaction history, which can lead to information loss and suboptimal predictions~\cite{chai2025longer}.

To mitigate this issue, memory-aware approaches leverage internal memory mechanisms to store and reuse user-specific hidden states\cite{lin2022sam,li2025vqlendtoendcontextawarevector}, thereby avoiding redundant computation. For example, MIMN~\cite{pi19prac} employs a Multi-Interest Memory Network with a Neural Turing Machine-style module to capture long-term user interests. More recently, MARM~\cite{lv2024marm} introduces a compute-storage separation architecture, caching intermediate representations to decouple online inference from the full sequence length, achieving notable improvements in short-video recommendation scenarios.

While these memory-based methods demonstrate promising efficiency gains, they often overlook the comparison of the computational overhead and storage costs associated with different memory management strategies. In this paper, we explicitly consider both aspects and reorganize existing approaches from the perspective of memory allocation granularity, which provides a more fine-grained and system-oriented taxonomy for benchmarking memory-aware long-sequence recommendation methods.
\section{Benchmark Design}
\definecolor{mygreen}{HTML}{9ac7bf}
\definecolor{myblue}{HTML}{a9c4eb}
\definecolor{myred}{HTML}{ea6b66}
\definecolor{myyellow}{HTML}{ffe599}
\definecolor{myorange}{HTML}{FFA500}

\tikzstyle{my-box}=[
    rectangle,
    draw=white,
    rounded corners,
    align=center,
    text opacity=1,
    minimum height=1.5em,
    minimum width=5em,
    inner sep=2pt,
    fill opacity=.8,
    line width=0.8pt,
]

\tikzstyle{leaf-head}=[my-box, minimum height=1.5em,
    draw=gray!80, %
    fill=gray!15,  %
    text=black, font=\normalsize,
    inner xsep=2pt,
    inner ysep=4pt,
    line width=0.8pt,
]

\tikzstyle{leaf-dataset}=[my-box, minimum height=1.5em,
    draw=mygreen, %
    fill=mygreen,  %
    text=black, font=\normalsize,
    inner xsep=2pt,
    inner ysep=4pt,
    line width=0.8pt,
]

\tikzstyle{leaf-fs}=[my-box, minimum height=1.5em,
    draw=myblue, %
    fill=myblue,  %
    text=black, font=\normalsize,
    inner xsep=2pt,
    inner ysep=4pt,
    line width=0.8pt,
]
\tikzstyle{leaf-bbm}=[my-box, minimum height=1.5em,
    draw=myred, %
    fill=myred,  %
    text=black, font=\normalsize,
    inner xsep=2pt,
    inner ysep=4pt,
    line width=0.8pt,
]
\tikzstyle{leaf-metrics}=[my-box, minimum height=1.5em,
    draw=myyellow, %
    fill=myyellow,  %
    text=black, font=\normalsize,
    inner xsep=2pt,
    inner ysep=4pt,
    line width=0.8pt,
]
\tikzstyle{leaf-architecture}=[my-box, minimum height=1.5em,
    draw=myorange, %
    fill=myorange,  %
    text=black, font=\normalsize,
    inner xsep=2pt,
    inner ysep=4pt,
    line width=0.8pt,
]

\tikzstyle{modelnode-dataset}=[my-box, minimum height=1.5em,
    draw=mygreen!70, %
    fill=mygreen!70,  %
    text=black, font=\normalsize,
    inner xsep=2pt,
    inner ysep=4pt,
    line width=0.8pt,
]

\tikzstyle{modelnode-architecture}=[my-box, minimum height=1.5em,
    draw=myorange!70, %
    fill=myorange!70,  %
    text=black, font=\normalsize,
    inner xsep=2pt,
    inner ysep=4pt,
    line width=0.8pt,
]

\tikzstyle{modelnode-fs}=[my-box, minimum height=1.5em,
    draw=myblue!70, %
    fill=myblue!70,  %
    text=black, font=\normalsize,
    inner xsep=2pt,
    inner ysep=4pt,
    line width=0.8pt,
]
\tikzstyle{modelnode-bbm}=[my-box, minimum height=1.5em,
    draw=myred!70, %
    fill=myred!70,  %
    text=black, font=\normalsize,
    inner xsep=2pt,
    inner ysep=4pt,
    line width=0.8pt,
]
\tikzstyle{modelnode-metrics}=[my-box, minimum height=1.5em,
    draw=myyellow!70, %
    fill=myyellow!70,  %
    text=black, font=\normalsize,
    inner xsep=2pt,
    inner ysep=4pt,
    line width=0.8pt,
]

\begin{figure}[]
    \centering
    \resizebox{1.0\linewidth}{!}{
        \begin{forest}
            forked edges,
            for tree={
                grow=east,
                reversed=true,
                anchor=base west,
                parent anchor=east,
                child anchor=west,
                base=left,
                font=\normalsize,
                rectangle,
                draw=white,
                rounded corners,
                align=left,
                minimum width=1em,
                edge+={darkgray, line width=1pt},
                s sep=5pt,
                l sep=20pt,
                inner xsep=3pt,
                inner ysep=3pt,
                line width=1pt,
                ver/.style={rotate=90, child anchor=north, parent anchor=south, anchor=center},
            },
            [%
                MALLOC,leaf-head
                [
                      Sequence Level \\Compression, leaf-dataset, text width=7em
                     [
                        Native, modelnode-dataset, text width=8em
                     ]                     
                     [
                        Linformer, modelnode-dataset, text width=8em
                     ]
                     [
                        Reformer, modelnode-dataset, text width=8em
                     ]
                ]
                [
                    Token Level \\Compression, leaf-fs, text width=7em
                    [
                        Merging, leaf-fs, text width=6em
                        [Longformer, modelnode-fs, text width=8em]
                        [Activation Beacon, modelnode-fs, text width=8em]
                    ]
                    [
                        Pruning, leaf-fs, text width=6em
                        [H2O, modelnode-fs, text width=8em]
                        [SnapKV, modelnode-fs, text width=8em]
                    ]
                ]
                [
                    Head Level \\Compression, leaf-bbm, text width=7em
                    [
                        MQA, modelnode-bbm, text width=8em
                    ]
                    [
                        GQA, modelnode-bbm, text width=8em
                    ]
                    [MLA, modelnode-bbm, text width=8em]
                ]
                [
                    Precision Level \\Compression, leaf-metrics, text width=7em
                    [
                        KIVI, modelnode-metrics, text width=8em
                    ]
                    [IntactKV, modelnode-metrics, text width=8em]
                ]
                [
                    Architecture \\Level, leaf-architecture, text width=7em
                    [
                        RWKV, modelnode-architecture, text width=8em
                    ]
                ]
            ]
        \end{forest}
    }
    \caption{Benchmark Overview.}
    \label{fig:overview}
    \vspace{-4mm}
\end{figure}
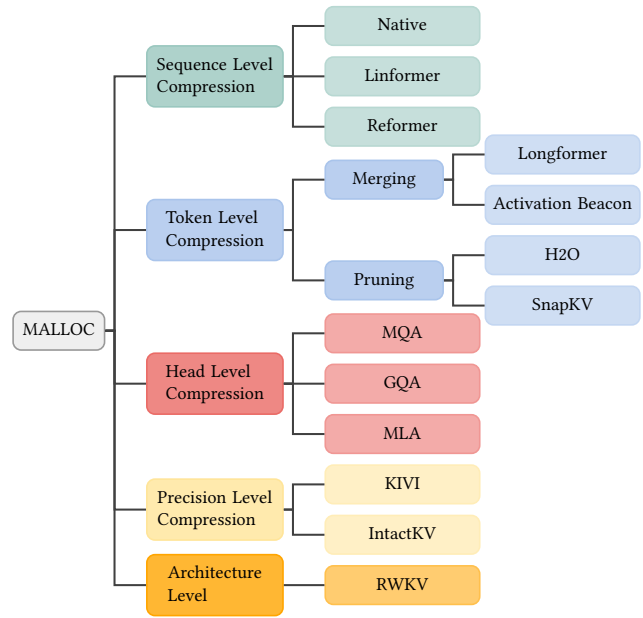
\begin{figure*}[ht]
    \centering
    \includegraphics[width=1.0\linewidth]{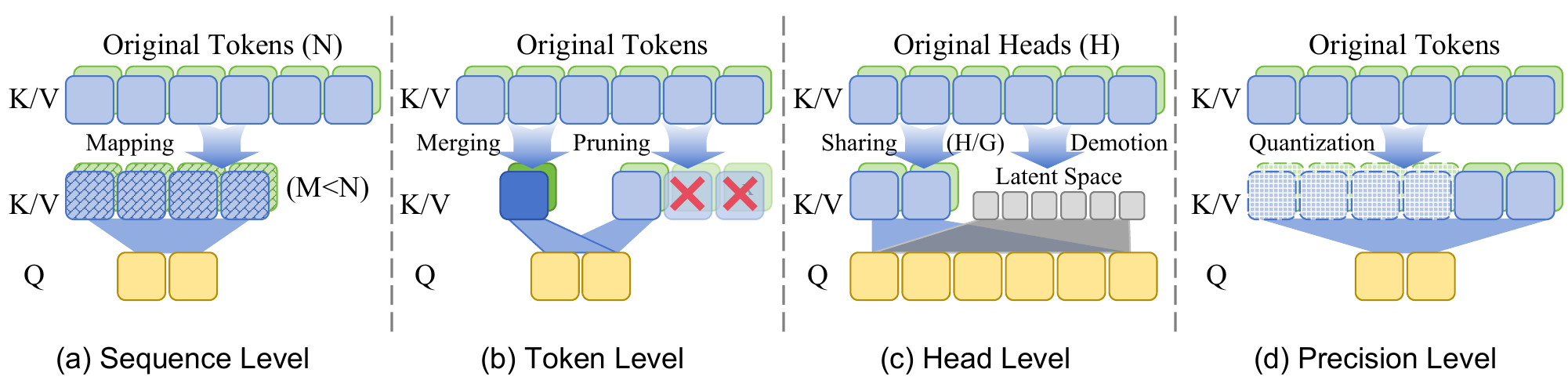}
\caption{A Visual Taxonomy of Multi-Level KV Cache Compression. (a) Transforms the full sequence of length $N$ into a compact representation of length $M$.
(b) Reduces the token count by either merging adjacent tokens or pruning less significant ones.
(c) Consolidates $H$ specific attention heads into $G$ shared or grouped KV heads to reduce memory redundancy.
(d) Compresses the storage bit-width of KV states from high-precision to low-precision grids without altering the sequence length.}
  \label{fig:show}
\end{figure*}
In this section, we will provide a more comprehensive description of our MALLOC mainly from the following five perspectives. 
We begin by laying out the fundamentals of large‐scale sequential recommendation in Section~\ref{sec:preliminary}, then introduce the memory‐aware baselines that anchor our evaluation in Section
~\ref{sec:baseline}. Section~\ref{sec:backbone} details the backbone architectures we employ, followed by a description of the real‐world datasets and how they interface with each backbone in Section~\ref{sec:datasets}. We conclude in Section~\ref{sec:metric} by defining the metrics used to assess performance in our experiments.

\subsection{Preliminary}
\label{sec:preliminary}

In practical large-scale deployments, the cardinalities of both the user set and the item set frequently reach billions of elements, thereby rendering scalability a primary design and methodological challenge. In recommender systems, the interaction space is typically characterized by two fundamental entity sets: a user set $\mathcal{U} = \{u_1, u_2, \ldots, u_{N_u}\}$ and an item set $\mathcal{V} = \{v_1, v_2, \ldots, v_{N_v}\}$. Here, $N_u$ and $N_v$ denote the total numbers of users and items, respectively. 

For each user $u \in \mathcal{U}$, we model their behavioral history as an interaction sequence $S_u = \{v_1, v_2, \ldots, v_{L_u}\}$, where $L_u$ is the length of the sequence and each $v_t \in S_u$ denotes the item from $\mathcal{V}$ with which the user interacted at discrete time step $t$.

Sequential recommendation aims to capture the temporal dynamics and evolving intent embedded in these sequences to accurately anticipate future actions. Specifically, given a user's historical sequence $S_u$, the objective is to estimate the likelihood that a candidate item $v \in \mathcal{V}$ will be clicked, which can be formulated as:
\begin{equation}
    \hat{y}_{u,v} = P(v \mid S_u, \Theta),
\end{equation}
where $\Theta$ denotes the model parameters learned during training and $\hat{y}_{u,v}$ represents the predicted probability. The prediction model is typically optimized via the standard cross-entropy loss over the training dataset $\mathcal{D}$:
\begin{equation}
    \mathcal{L} = - \frac{1}{|\mathcal{D}|} \sum_{(u, v) \in \mathcal{D}} \left( y_{u,v} \log \hat{y}_{u,v} + (1 - y_{u,v}) \log (1 - \hat{y}_{u,v}) \right).
\end{equation}
where $y_{u,v} \in \{0, 1\}$ indicates the ground truth label, representing whether user $u$ actually engaged with the candidate item $v$.

\subsection{Baselines}
\label{sec:baseline}
Compression techniques for large-scale sequential recommendation models can be categorized from several angles. One common perspective is based on \textit{scope}: while many methods focus on reducing memory consumption to accelerate inference, others target the speed-up of specific model components without explicitly optimizing memory usage across the full inference path. Another perspective centers on \textit{model architecture}, where compression is achieved through architectural changes such as lightweight layers or sparse connections.
However, these perspectives remain relatively coarse and often overlook a more nuanced understanding of how memory is allocated and utilized. 
In this work, we organize existing approaches primarily according to the \textbf{granularity of memory allocation}, which subsumes and refines prior categorization strategies.
Specifically, as illustrated in Figure~\ref{fig:overview}, we classify memory-aware compression methods into five categories.
This finer-grained view enables a more systematic analysis of memory-centric design choices across different compression paradigms.

\subsubsection{Sequence-Level Compression}
Sequence-level compression methods reduce computational complexity by approximating or restructuring attention along the sequence dimension, while largely preserving the original token representations. These approaches do not modify the internal structure of tokens or attention heads, but instead limit how tokens interact across long sequences.
\begin{itemize}
    \item \textbf{Native}~\cite{vaswani2017attention} approach does not incorporate any additional compression mechanisms; it simply leverages cached historical states when processing user requests, reducing the quadratic complexity with respect to sequence length to linear time, at the cost of substantial memory occupation.
    \item \textbf{Linformer}~\cite{wang2020linformer} maintains the full cache of hidden states yet projects keys and values into a lower-dimensional space using fixed linear mappings. This low-rank approximation reduces both computation and intermediate memory for attention while still storing all original hidden states.
    \item \textbf{Reformer}~\cite{Kitaev2020Reformer} introduces a hashing strategy that only computes attention with tokens that share the same hash value to accelerate computation without any memory management.
\end{itemize}
\begin{table*}[ht]
\centering
\setlength{\tabcolsep}{7pt}
\renewcommand{\arraystretch}{0.95}
\caption{Statistics of the datasets.}
\label{tab:dataset}\resizebox{1.0\textwidth}{!}{
\begin{tabular}{
    l
    l
    S[table-format=6]
    S[table-format=7]   % 修正为 7 位数，因为 1704882 是 7 位
    S[table-format=7]   % 13,135,234 是 7 位整数 + 可能的逗号
    S[table-format=4]
    c % 最大长度 1024 是三位整数
    S[table-format=1.2]
}
\toprule
Dataset & Domain & {\#Users} & {\#Items} & {\#Interactions} & {Max Length} & {Avg Length} & {Positive Rate} \\
\midrule
Amazon-Electronic & E-commerce & 192403 & 63003 & 3333478 & 128 & \ 9.00 & 0.44 \\
MicroVideo1.7M    & Micro-videos & 10988 & 1704882 & 13135234 & 1024 & 502 & 0.15 \\
KuaiVideo           & Micro-videos & 10002 & 632406 & 10869055 & 1024 & 416 & 0.17 \\
\bottomrule
\end{tabular}}
\end{table*}
\subsubsection{Token-Level Compression}
Token-level compression methods operate at the granularity of individual tokens by merging, pruning, or selectively retaining intermediate states. By reducing the number of active tokens participating in attention, these approaches directly lower both memory consumption and computational cost, making them particularly effective for long-sequence scenarios.
\begin{itemize}
    \item \textbf{Merging} \begin{itemize}
    \item Longformer~\cite{Beltagy2020Longformer} 
    employs a structured attention pattern that combines local sliding-window attention with a small set of global tokens. This design effectively limits token-level interactions and reduces the number of active token states involved in attention computation.
        \item Activation Beacon~\cite{zhanglong} 
        introduces a compact set of learnable beacon tokens that aggregate information from long sequences. By distilling historical interactions into a compact representation, this method reduces the number of stored and attended tokens while preserving long-term contextual information, yielding a more efficient inference process without sacrificing representation quality.
    \end{itemize}
    \item \textbf{Pruning} \begin{itemize}
        \item H2O~\cite{zhang2023h2o} observed that only a small fraction of tokens have a substantial impact on the final output when computing attention. As a result, it retains only the intermediate states of these critical tokens while discarding the rest.
        \item SnapKV~\cite{li2024snapkv} 
        exploits the stability of attention patterns across decoding steps, compressing the KV cache by preserving only persistently attended tokens. This selective retention achieves efficient memory usage with minimal structural modification to the model.
    \end{itemize}
\end{itemize}

\subsubsection{Head-Level Compression} 
Head-level compression methods reduce memory usage by sharing or jointly compressing attention heads within the same layer. Instead of reducing the number of tokens, these approaches lower the dimensionality or redundancy of attention representations across heads.
\begin{itemize}
    \item \textbf{MQA}~\cite{shazeer2019fast} 
    shares a single set of key and value vectors across all attention heads, reducing both the computational cost and memory usage, thereby improving the inference efficiency.
    \item \textbf{GQA}~\cite{ainslie2023gqa} improves inference efficiency by grouping and sharing keys/values across multiple attention heads. MQA is a special case of this mechanism with only one group.
    \item \textbf{MLA}~\cite{liu2024deepseek} 
    applies low-rank joint compression over attention heads, where the KV cache during inference is reduced, significantly lowering memory consumption while maintaining model performance.
\end{itemize}

\subsubsection{Precision-Level Compression} 
Precision-level methods provide a lightweight memory preservation strategy by lowering the numerical precision of cached intermediate states, without modifying the model architecture or attention structure.
\begin{itemize}
    \item \textbf{KIVI}~\cite{liu2024kivi} 
    applies asymmetric low-bit quantization to the KV cache, leveraging a 2-bit KV cache quantization approach that operates without the need for fine-tuning.
    \item \textbf{IntactKV}~\cite{liu2024intactkv} 
    preserves full-precision representations for critical tokens while quantizing others, mitigating performance degradation caused by aggressive quantization.
\end{itemize}

\subsubsection{Architecture Level} 
Architecture level fundamentally alters the sequence modeling paradigm such that the current hidden state depends only on a compact recurrent state, eliminating the need to store long token histories.
\begin{itemize}
    \item \textbf{RWKV}~\cite{peng2023rwkv} is a novel sequence modeling architecture that combines the strengths of Recurrent Neural Networks and Transformer attention mechanisms. By introducing the gating mechanism, it retains the efficiency of RNNs while achieving capabilities similar to those of Transformers.
\end{itemize}

\subsection{Backbone}
\label{sec:backbone}
In recent industrial systems such as MTGR~\cite{han2025mtgr} and LONGER~\cite{chai2025longer}, substantial gains have been achieved by integrating rich item features alongside sequential data under conventional training schemes. In contrast, MALLOC is designed to explore how pure sequence–based recommendation models scale as their depth and capacity grow. To establish a consistent benchmark, we adopt HSTU~\cite{zhai2024actions} as our backbone. HSTU extends the standard Transformer by tailoring its attention layers to capture the evolving, non-stationary patterns in user behavior found in large recommendation logs. This specialized design both preserves the model’s ability to handle characteristics inherent to long sequence datasets and delivers notable speedups compared to traditional Transformers.

\subsection{Datasets}
\label{sec:datasets}
In order to thoroughly demonstrate the validity of our benchmark, we assess existing methods across three widely used real-world datasets shown in Table~\ref{tab:dataset}. Owing to the relatively sparse characteristics and popularity in research, Amazon-Electronic is chosen to investigate the behavior of baseline methods in the context of sparse and short interaction sequences ($\le 150$). To explore the long sequence modeling, MALLOC adopts MicroVideo1.7M with more sequential data, and the interaction of each user within it could be up to 1000. Additionally, to compare the performance of different methods in scenarios closer to real-world recommender systems, KuaiVideo, an industrial dataset from Kuaishou, is further selected to provide longer user-item interactions. 

\subsection{Metrics}
\label{sec:metric}
In this paper, apart from the ranking metrics widely adopted in mainstream recommendation systems, we further incorporate a set of resource-aware indicators that are particularly meaningful for practical online deployment under resource constraints, such as the induced storage overhead and computational cost. Specifically, MALLOC introduces three metrics, AUC, GAUC, and Logloss for recommendation evaluation, and another two metrics, MACS, and memory occupation for resource calculation.

\begin{itemize}
\item \textbf{AUC} (Area Under the ROC Curve) measures the probability that a positive instance is ranked higher than a negative one. It is a standard metric to evaluate the ranking capability of recommenders. The formula is defined as:
\begin{equation}
    \mathrm{AUC} = \frac{1}{|\mathcal{P}|} \sum_{(u, v^+, v^-) \in \mathcal{P}} \mathbb{I}(\hat{y}_{u, v^+} > \hat{y}_{u, v^-}),
\end{equation}
where $\mathcal{P}$ denotes the set of all valid triplets $(u, v^+, v^-)$ in the test set, composed of a user $u$, a positive item $v^+$, and a negative item $v^-$. $\hat{y}_{u, v^+}$ and $\hat{y}_{u, v^-}$ represent the predicted scores for the positive and negative items, respectively, and $\mathbb{I}(\cdot)$ is the indicator function that returns 1 if the condition holds and 0 otherwise.

\item \textbf{GAUC} (Grouped AUC) extends AUC by computing it separately for each user and then calculating the weighted average across all users. This metric is particularly useful in recommendation scenarios where user activity levels vary significantly. For the user set $\mathcal{U}$, the formula is:
\begin{equation}
    \mathrm{GAUC} = \frac{\sum_{u \in \mathcal{U}} N_u \cdot \mathrm{AUC}_u}{\sum_{u \in \mathcal{U}} N_u},
\end{equation}
where $N_u$ denotes the number of impressions for user $u$, and $\mathrm{AUC}_u$ is the AUC calculated specifically on the user $u$.

\item \textbf{Logloss} (Logarithmic Loss) quantifies the discrepancy between the predicted interaction probability and the ground truth label. It is widely used as the optimization objective during training to instruct the update of model parameters.
  \item \textbf{MACs} (Multiply–Accumulate Operations) represent the number of elementary multiply-and-add operations performed by a model during inference, which focuses specifically on the core computational workload.
  \item \textbf{Memory Occupation} measures the amount of memory for storing the hidden states. This is especially important when the acceleration of MALLOC is brought about by a larger dependency on storage space.
\end{itemize}

\section{Experiments}
\begin{table*}[htb]
\centering
\caption{Overall Performance of different memory-aware methods. In this table, we report AUC ($\Uparrow$), GAUC ($\Uparrow$), and Logloss ($\Downarrow$), three metrics for recommendation. The best results are \textbf{bolded}, and the second-best results are \underline{underlined}.}
\label{tab:experiment_rec}
\setlength{\tabcolsep}{7pt}
\renewcommand{\arraystretch}{0.95}
\resizebox{1.0\textwidth}{!}{
\begin{tabular}{c|c|ccc|ccc|ccc}
\toprule
\multirow{2}{*}{\textbf{Type}} & \multirow{2}{*}{\textbf{Baseline}} & 
\multicolumn{3}{c|}{\textbf{MicroVideo}} & 
\multicolumn{3}{c|}{\textbf{KuaiVideo}} & 
\multicolumn{3}{c}{\textbf{Amazon}} \\
\cmidrule(lr){3-5} \cmidrule(lr){6-8} \cmidrule(l){9-11}
 & & \textbf{AUC} & \textbf{GAUC} & \textbf{Logloss} & 
 \textbf{AUC} & \textbf{GAUC} & \textbf{Logloss} & 
 \textbf{AUC} & \textbf{GAUC} & \textbf{Logloss} \\
\midrule
\multirow{3}{*}{Sequence Level} & Native & 0.763 & 0.704 & 0.396 & 0.776 & 0.691 & 0.417 & 0.931 & 0.932 & 0.378 \\
                             & Linformer & 0.735 & 0.686 & 0.415 & 0.746 & 0.667 & 0.438 & \textbf{0.964} & \textbf{0.987} & \textbf{0.166} \\
                             & Reformer & 0.769 & 0.711 & 0.392 & 0.787 & 0.705 & 0.410 & \underline{0.947} & \underline{0.945} & \underline{0.372} \\
\midrule
\multirow{2}{*}{Merging} & Longformer & \underline{0.835} & \textbf{0.806} & \underline{0.345} & \underline{0.840} & \underline{0.707} & \underline{0.366} & 0.901	& 0.876	& 0.439 \\
& Beacon & \textbf{0.836} & \underline{0.714} & \textbf{0.228} & \textbf{0.864} & \textbf{0.711} & \textbf{0.209} & 0.926 & 0.906 & 0.440 \\
\midrule
\multirow{2}{*}{Pruning}     & H2O & 0.686 & 0.660 & 0.474 & 0.669 & 0.647 & 0.552 & 0.881 & 0.879 & 0.483 \\
                             & SnapKV & 0.708 & 0.668 & 0.447 & 0.694 & 0.649 & 0.533 & 0.881 & 0.879 & 0.482 \\
\midrule
\multirow{3}{*}{Head Level}     & MQA & 0.762 & 0.703 & 0.396 & 0.775 & 0.690 & 0.417 & 0.929 & 0.930 & 0.393 \\
                             & GQA & 0.762 & 0.704 & 0.396 & 0.775 & 0.690 & 0.417 & 0.932 & 0.932 & 0.382 \\
                             & MLA & 0.762 & 0.704 & 0.396 & 0.776 & 0.691 & 0.417 & 0.932 & 0.931 & 0.391 \\
\midrule
\multirow{2}{*}{Precision Level} & KIVI & 0.724 & 0.681 & 0.436 & 0.679 & 0.649 & 0.583 & 0.888 & 0.883 & 0.468 \\
                              & IntactKV & 0.664 & 0.649 & 0.644 & 0.665 & 0.644 & 0.729 & 0.886 & 0.882 & 0.476 \\
\midrule
Architecture & RWKV & 0.765 & 0.705 & 0.393 & 0.778 & 0.692 & 0.416 & 0.893 & 0.895 & 0.455 \\
\bottomrule
\end{tabular}}
\end{table*}

In this section, we present a comprehensive evaluation of the MALLOC benchmark. To systematically analyze the effectiveness and efficiency of memory-aware long sequence compression, our experiments are designed to answer the following four key research questions.
\textbf{RQ1:} How do different memory management strategies impact the recommendation accuracy?
\textbf{RQ2:} To what extent can these methods reduce the memory footprint and computational overhead during inference? 
\textbf{RQ3:} How do these methods balance the conflict between recommendation performance and resource consumption? 
\textbf{RQ4:} What are the implications of these methods regarding implementation complexity and scalability when applied to large-scale industrial models? 

\subsection{Data Processing}
In this paper, based on the sparsity and average length of each dataset, we set 128, 1024, and 1024 for Amazon Electronics, MicroVideo, and KuaiVideo, respectively. Since we adopt HSTU, a backbone model already widely used in the industry as our foundation, the input sequence not only contains item IDs but also includes their corresponding click labels. Therefore, MALLOC chooses to concatenate all the labels from the dataset in chronological order to form a new sequence as part of the HSTU input, thereby preserving the click label information. 

\subsection{Experimental Details}
We implement this benchmark based on FuxiCTR~\cite{zhu2021open} using PyTorch 2.6.0, an open-source framework for CTR prediction. All evaluated methods are implemented under a unified codebase and backbone to ensure fair comparison. Unless otherwise specified, the same set of hyperparameters is used across all methods without per-method tuning.
The learning rate $lr$ is fixed to $5\times10^{-4}$ for all datasets. The item embedding dimension is set to 256, and each model employs 8 attention heads. All reported results are averaged over three independent runs with different random seeds.

All resource measurements were conducted on a single machine running Ubuntu 4.15.0-208-generic, equipped with an NVIDIA GeForce RTX 3090 GPU (24GB) and 10 Intel Xeon Gold 6240 CPUs. MACs are computed using the \textbf{ptflops} package, which counts operations from dense linear layers and attention projections; embedding lookups and non-linear operations are excluded. Memory occupation reports the peak KV cache memory per inference request, measured during inference with batch size set to 8.

\subsection{Overall Performance (RQ1)}
To rigorously assess the validity of our benchmark evaluation, we perform experiments using three distinct random seeds across multiple representative baseline methods on three real-world datasets. In Table~\ref{tab:experiment_rec}, we report the variability in recommendation performance induced by different memory management, which constitutes a primary factor of interest in this research area. The empirical results further indicate that modifications in memory allocation and management strategies can yield substantial discrepancies in the final performance outcomes.

In the \textit{Sequence Level} category, the \textbf{Reformer}, based on adaptive type selection, identifies the most relevant historical states for computation through local hashing based on candidate items. This approach not only reduces the state information transferred to GPU memory but also ensures the scoring scale is determined solely by relevant items, a strategy proven effective in recommendation systems~\cite{si2024twin,chen2021end}. Consequently, Reformer achieves significant improvements over Native across all three datasets. Conversely, \textbf{Linformer} exhibits a distinct anomaly: it outperforms other baselines on the short-sequence Amazon dataset but underperforms on the long-sequence KuaiVideo and MicroVideo. Such dataset-specific gains likely stem from compression along timesteps interacting with label-concatenated inputs, which leads to data leakage during training. In this case, the model may copy future patterns rather than inferring behavior from historical interactions.

\textit{Merging} methods process sequences in segments or aggregates to reduce length. \textbf{Longformer} consistently achieves the best or second-best results in terms of AUC and GAUC on long-sequence datasets (MicroVideo and KuaiVideo), demonstrating the effectiveness of localized attention in modeling long-range dependencies. However, on the Amazon dataset, this trend reverses. This aligns with the fact that under short-sequence settings, Longformer's attention window nearly covers the entire sequence, diminishing its structural advantage and introducing unnecessary computation. Similarly, \textbf{Activation Beacon} consistently achieves strong performance on long-sequence datasets, with notable improvements in Logloss. This indicates that selectively aggregating historical representations improves prediction calibration while preserving discriminative power. However, its gains in GAUC are generally more modest than in Logloss, implying that memory merging primarily stabilizes instance-level predictions rather than substantially reshaping user-level ranking.
In contrast, \textit{Pruning} methods such as \textbf{H2O} and \textbf{SnapKV} maintain stable but relatively lower performance across all datasets. Their consistent yet modest results imply that simple retention or eviction strategies alone are insufficient to preserve critical historical information for recommendation accuracy, especially in long-sequence scenarios.

In the \textit{Head Level} category, approaches including \textbf{MLA}, \textbf{MQA}, and \textbf{GQA} deliver performance comparable to the Native baseline across all datasets. While these methods strike a favorable balance between efficiency and accuracy, they generally fail to surpass full-memory or merging-based techniques on long sequences. This outcome highlights the inherent limitations of parameter-sharing mechanisms in capturing the diverse and complex user behaviors embedded within extended interaction histories.

\textit{Precision Level} methods, specifically \textbf{KIVI} and \textbf{IntactKV}, suffer from marked performance degradation, particularly on long-sequence datasets. This decline suggests that the aggressive quantization of cached states compromises the fidelity of historical information, which is indispensable for accurately modeling long-term user intent in recommendation tasks.

\textbf{RWKV} diverges from the standard backbone by employing convolutional operations to process input sequences in parallel. Although it yields competitive results in dense scenarios, it struggles with sparse datasets. This indicates that its coarser modeling of sequential dependencies may lack the granularity required to capture subtle signals in sparse interaction data.

Overall, the results indicate that memory-aware designs do not universally improve recommendation accuracy, but their effectiveness is highly contingent on how historical information is summarized, retained, and reused. While some methods excel in pure effectiveness under specific data regimes, others offer more balanced performance when resource constraints are considered. This observation motivates the subsequent trade-off analysis, where we further examine how these methods balance effectiveness against computational and memory costs.
\begin{figure*}[ht]
    \centering
    \includegraphics[width=0.94\linewidth]{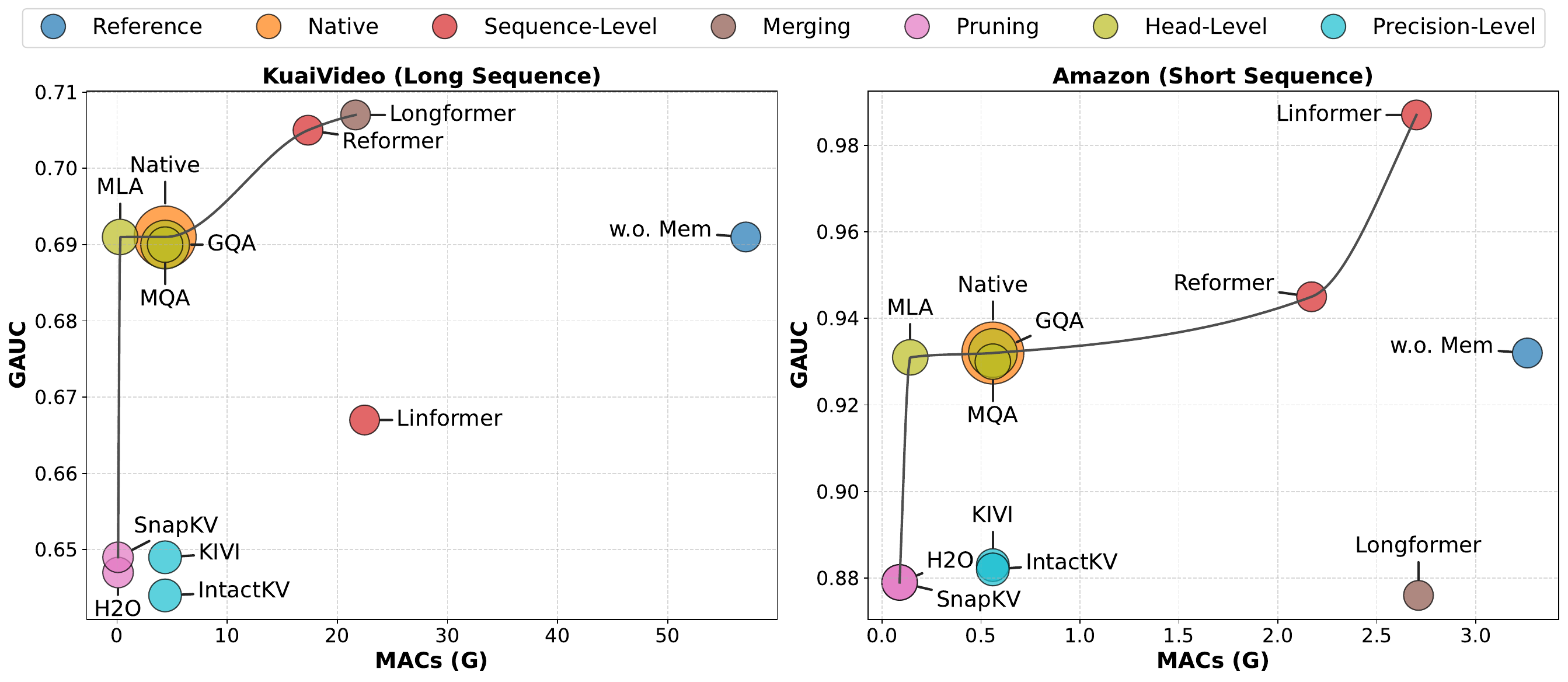}
\caption{The Trade-off Landscape between Recommendation Performance and Resource Consumption. 
  The visualization contrasts GAUC against computational cost (MACs) on KuaiVideo (Left) and Amazon (Right) datasets. 
  The bubble size is proportional to the memory occupation.}
  \label{fig:Pareto}
  \vspace{-1em}
\end{figure*}
\begin{table}[tb]
\centering
\caption{Analysis of the Resource Constraint.}
\label{tab:experiment_resource}
\begin{tabular}{c|cc|cc}
\toprule
\multirow{2}{*}{\textbf{Baseline}} & \multicolumn{2}{c|}{\textbf{KuaiVideo}} & \multicolumn{2}{c}{\textbf{Amazon}} \\
\cmidrule(lr){2-3} \cmidrule(lr){4-5}
                                   & \textbf{MACs} & \textbf{Memory} & \textbf{MACs} & \textbf{Memory} \\
\midrule
w.o. Mem                        & 57.1 G             & 0.0               &3.26 G             & 0.0               \\
Native                          & 4.36 G             & 127.88 MB                & 560.16 M             & 15.88 MB               \\
Linformer                       &   22.48G           & 0.0   & 2.70G    & 0.0               \\
Reformer                        &  17.34G           & 0.0  & 2.17G   & 0.0           \\
Longformer                      &  21.66G           & 0.0  & 2.71G   & 0.0           \\
H2O                             & 89.37 M             & 2 MB                & 89.37 M           & 2 MB               \\
SnapKV                          & 89.37 M             & 2 MB                & 89.36 M           & 2 MB              \\
MQA                             & 4.36 G             & 15.98 MB                & 560.16 M             & 1.98 MB               \\
GQA                             & 4.36 G             & 63.94 MB                & 560.16 M             & 7.94 MB               \\
MLA                             & 276.25 M             & 15.98 MB                & 143.0 M             & 1.98 MB               \\
KIVI                            & 4.36 G             & 7.99 MB                & 560.16 M           & 0.99 MB               \\
IntactKV                        & 4.36 G             & 7.99 MB                & 560.16 M            & 0.99 MB               \\
\bottomrule
\end{tabular}
\vspace{-1.2em}
\end{table}
\subsection{Analysis of Resource Consumption (RQ2)}
In real-world scenarios, besides the model's recommendation capability, efficiently deploying the model to handle requests from hundreds of millions of users in a short time presents an even greater challenge. Table \ref{tab:experiment_resource} reports the MACs and memory occupation of representative baselines on both long-sequence (KuaiVideo) and short-sequence (Amazon) datasets. These results highlight how different design paradigms trade computation for memory under varying sequence lengths.

Native only needs to perform target attention based on the current behavior during autoregressive prediction, avoiding the quadratic self-attention computation with respect to sequence length; however, resulting in substantial memory occupation. Although this per-user memory cost may appear moderate in isolation, it becomes prohibitive in large-scale recommendation systems with billions of users and persistent historical storage. This observation motivates the need for memory-aware designs that explicitly balance computation and storage.

Methods without memory usage require computing attention over the entire sequence for each request. Although they incur no memory overhead, they introduce significant computational complexity. Reformer, Linformer, and Longformer reduce the time complexity of the attention computation by employing techniques such as locality-sensitive hashing, linear projections along the time dimension, and sliding-windows attention mechanism, enabling each token to interact with only a small subset of other tokens during the inner product calculation.

MQA and GQA reduce the required memory size by sharing the same intermediate states across multiple attention heads. MLA performs dimensionality reduction on the token level and then upscales after computation, which simultaneously reduces both the number of floating-point operations and the amount of stored states. Their consistent resource profiles indicate that computation and memory costs are largely determined by architectural choices rather than implementation details.

Among token-level compression methods, merging-based approaches such as Activation Beacon substantially reduce the number of stored intermediate states by aggregating multiple tokens into compact representations.
Similarly, pruning-based approaches like H2O and SnapKV permanently remove less important tokens during the prefilling phase, leading to substantial acceleration in the autoregressive generation stage. 
In contrast, quantization-based methods 
retain the same computational cost as Native but significantly reduce memory occupation, illustrating a complementary optimization direction that prioritizes storage efficiency over computation reduction.
aim to compress the model by reducing the precision of stored states after training. However, since we retain the network parameters in full precision, these methods still incur high computational costs during inference.

\subsection{Accuracy–Resource Pareto Frontier (RQ3)}
\begin{figure*}[ht]
    \centering
    \includegraphics[width=1\linewidth]{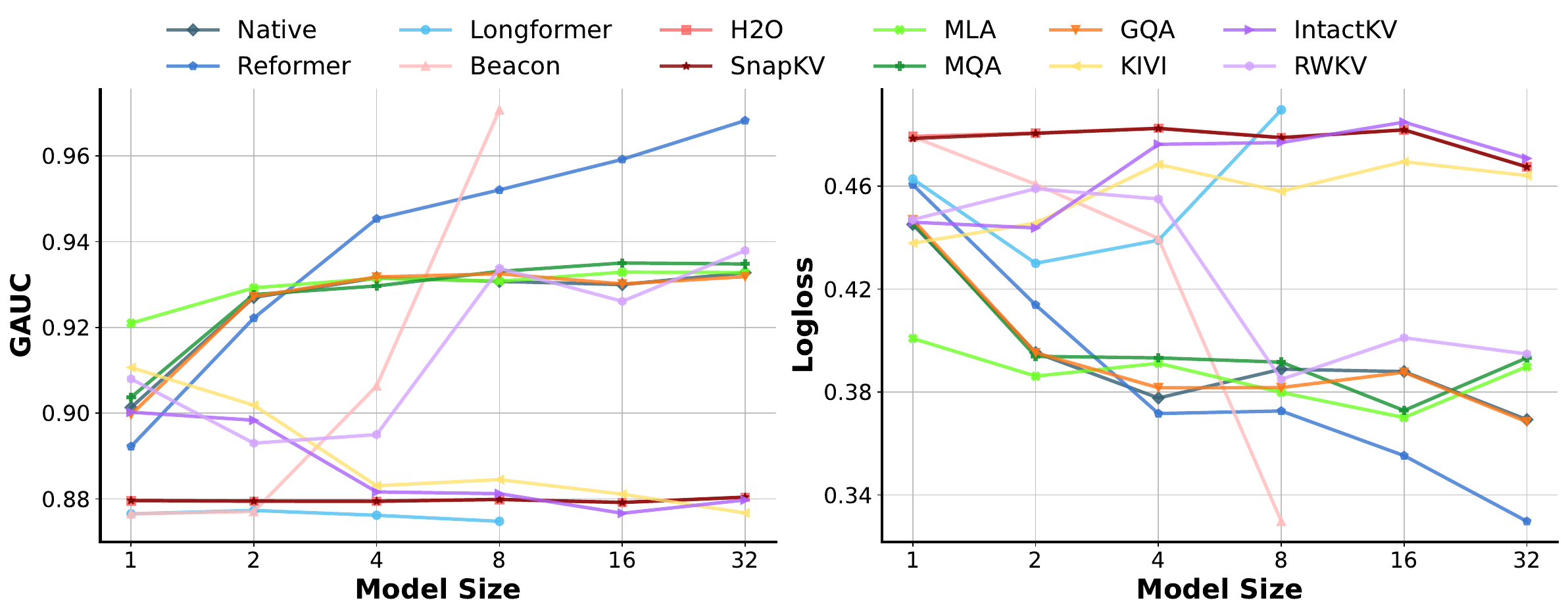}
    \caption{Scalability Analysis on Amazon Dataset. The recommendation performance (GAUC, Logloss) of different memory-aware methods as the model depth increases from 1 to 32 blocks.}
    \label{fig:placeholder}
\end{figure*}
As shown in Figure~\ref{fig:Pareto}, to highlight how recommendation accuracy varies jointly with computational and memory budgets instead of focusing on a single metric, we compare different memory-aware approaches from the perspective of accuracy–resource trade-offs on KuaiVideo and Amazon. This perspective shifts the focus from a single metric to a joint assessment of GAUC against computational cost (MACs) and memory occupation.

\textbf{High-Resource, High-Accuracy Regime.} \textit{Sequence Level} methods, particularly \textbf{Native} and \textbf{Reformer}, occupy the high-resource end of the frontier. While they generally achieve superior GAUC by retaining full context fidelity, they are associated with prohibitive computational costs or substantial memory occupation (as shown in Table~\ref{tab:experiment_resource}, Reformer requires up to 17.34 G MACs on KuaiVideo). This makes them less ideal for strictly constrained industrial environments despite their predictive dominance.

\textbf{Low-Resource, Efficiency-First Regime.} Conversely, \textit{Token Level (Pruning)} methods, along with \textit{Precision Level} approaches, form a distinct cluster characterized by minimal resource consumption. As detailed in Table~\ref{tab:experiment_resource}, these methods reduce memory usage to as low as $\approx$2 MB. However, this aggressive compression comes at a cost: their GAUC is comparatively lower, especially on the long-sequence benchmark. This finding suggests that simply discarding or quantizing tokens limits the model's ability to retain long-range behavioral dependencies critical for sequential recommendation.

\textbf{Balanced Regime.} \textit{Head Level} compression methods, including \textbf{MQA}, \textbf{GQA}, and \textbf{MLA}, occupy an intermediate position. They realize a more favorable trade-off by providing predictive performance comparable to Native (Native-level GAUC) while significantly constraining both computational complexity and memory consumption (reducing MACs from 4.36 G to $\approx$276 M). Notably, these methods emerge as non-dominated solutions on the Pareto frontier, offering competitive operating points for scenarios requiring a balance between accuracy and deployment costs.

In summary, Figure~\ref{fig:Pareto} demonstrates that accuracy improvements are inseparable from resource considerations. The absence of a universally dominant method underscores the necessity of the MALLOC benchmark, enabling practitioners to select appropriate memory-aware designs based on specific deployment constraints.

\subsection{Analysis of the Scalability (RQ4)}
To better understand how compression and acceleration techniques affect model scalability, we evaluate their performance and training stability across varying model sizes using the Amazon dataset. 

As shown in Figure~\ref{fig:placeholder}, including the Native, Reformer, GQA, MQA, and MLA, increasing the number of blocks from shallow to moderate depth yields noticeable performance improvements. This trend indicates that deeper models can better exploit long-range sequential dependencies when the memory mechanism effectively preserves contextual information. However, the performance gains are not strictly monotonic. In multiple cases, accuracy saturates or fluctuates as depth further increases, suggesting diminishing returns beyond an architecture-specific depth.

Scalability in practice is also constrained by optimization stability. Certain architectures, such as Longformer and Beacon, fail to converge when the number of blocks exceeds moderate depth, exhibiting gradient explosion during training. This behavior reveals that computationally efficient long-sequence designs do not necessarily translate into stable deep architectures in recommendation settings. In contrast, Reformer and grouped-attention methods remain trainable across a broader range of depths, demonstrating stronger robustness to deep stacking.

Pruning methods show stable but nearly depth-invariant performance across all tested block numbers. While these methods scale favorably in terms of stability, the lack of performance improvement suggests that additional layers provide limited representational benefit, likely due to aggressive information reduction in memory.
Similarly, low-precision methods exhibit diminishing or negative returns as depth increases, indicating that approximation errors may accumulate across layers and hinder effective scaling.

Overall, our results demonstrate that memory-aware methods exhibit heterogeneous scalability behaviors. Optimal performance is typically achieved at intermediate depths, while excessive depth may lead to instability or ineffective context utilization. These findings emphasize that scalability in large sequential recommendation models requires a careful balance between architectural depth and memory design, rather than naive depth scaling.

\newcommand{\yes}{\ding{52}} % ✅
\newcommand{\no}{\ding{56}}   % ❌
\begin{table}[tb]
    \centering
    \caption{Comparison of engineering cost and workflow time for all the baselines. We use \yes$\star$, \yes, and \no to indicate the most suitable, moderately suitable, and least suitable methods for this domain, respectively.}
    \label{tab:baseline_comparison}
    \begin{tabular}{lcccc}
        \toprule
        \multirow{2}{*}{\textbf{Baseline}} & \multicolumn{2}{c}{\textbf{Engineering Cost}} & \multicolumn{2}{c}{\textbf{Workflow}} \\
        \cmidrule(lr){2-3} \cmidrule(lr){4-5}
        & \textbf{Easy} & \textbf{Hard} & \textbf{Convenient} & \textbf{Complex} \\
        \midrule
        Native            & \yes$\star$ & \no\,\,\,  & \yes$\star$ & \no\,\,\,  \\
        Linformer         & \yes$\star$ & \no\,\,\,  & \no\,\,\,  & \yes\,\,\, \\
        Reformer          & \yes\,\,\,  & \no\,\,\, & \no\,\,\,  & \yes\,\,\, \\
        Longformer        & \yes\,\,\,  & \no\,\,\, & \no\,\,\,  & \yes\,\,\, \\
        Beacon            & \no\,\,\,  & \yes\,\,\, & \no\,\,\,  & \yes\,\,\, \\
        H2O               & \yes\,\,\,  & \no\,\,\, & \yes\,\,\, & \no\,\,\,  \\
        SnapKV            & \yes\,\,\,  & \no\,\,\, & \yes\,\,\, & \no\,\,\,  \\
        MQA               & \yes$\star$ & \no\,\,\,  & \no\,\,\,  & \yes\,\,\, \\
        GQA               & \yes$\star$ & \no\,\,\,  & \no\,\,\,  & \yes\,\,\, \\
        MLA               & \yes$\star$ & \no\,\,\,  & \no\,\,\,  & \yes\,\,\, \\
        KIVI              & \yes\,\,\,  & \no\,\,\, & \yes\,\,\, & \no\,\,\,  \\
        IntactKV          & \yes\,\,\,  & \no\,\,\, & \yes\,\,\, & \no\,\,\,  \\
        RWKV              & \no\,\,\,  & \yes\,\,\, & \no\,\,\,  & \yes\,\,\, \\
        \bottomrule
    \end{tabular}
\end{table}
\subsection{Analysis of the Implementation (RQ4)}
In this section, we analyze the implementation costs of the considered methods, encompassing both the \textbf{implementation difficulty} and the \textbf{workflow complexity}, as summarized in Table~\ref{tab:baseline_comparison}.

\textbf{1)} From a development-cost perspective, baselines such as MQA and MHA require only minor code modifications, rendering them comparatively straightforward to implement. By contrast, approaches based on weight merging or quantization generally necessitate extensive alterations to the attention mechanism, thereby increasing both implementation complexity and engineering overhead. Moreover, architectural innovations in models such as RWKV entail substantial redesign and re-engineering, which pose significant challenges for deployment in industrial settings.

\textbf{2)} Some methods like H2O, SnapKV, and KIVI enable efficient
inference acceleration by leveraging a pre-trained large-scale recommendation model without requiring any additional training. In
contrast, other approaches necessitate modifying the model during training to incorporate compression-aware mechanisms, which
introduces a more complex workflow and pretraining costs.

\section{Conclusion}
In this work, we introduce MALLOC, the first comprehensive benchmark that systematically restructures the landscape of long-sequence compression through a novel taxonomy centered on memory allocation granularity. By categorizing existing compression techniques within a unified framework and evaluating them along the dimensions of predictive accuracy, computational cost, memory utilization, and scalability, we deliver a comprehensive view of the practical trade-offs underlying long-sequence recommendation. Our empirical analysis demonstrates that memory-aware acceleration is not a single, monolithic design decision, but rather a structured design space in which distinct compression granularities induce fundamentally different system and modeling behaviors. 
Beyond performance comparison, MALLOC functions as a diagnostic instrument for understanding where and why existing approaches succeed or fail. 
We argue that MALLOC constitutes an essential intermediate milestone for the research and practitioner community. It not only offers practitioners concrete guidelines for addressing the "Memory–Latency Dilemma" in industrial deployments, but also directs future research toward hardware–algorithm co-design and adaptive compression strategies that dynamically balance contextual information retention against system-level constraints.

\balance
\bibliographystyle{ACM-Reference-Format}
\bibliography{MALLOC}

\end{document}